\documentclass[]{IEEEtran}

\input epsf
\usepackage{graphicx}
\usepackage{subcaption}

\usepackage{url}

\usepackage{amsmath}   % From the American Mathematical Society
\DeclareRobustCommand*{\IEEEauthorrefmark}[1]{%
  \raisebox{0pt}[0pt][0pt]{\textsuperscript{\footnotesize\ensuremath{#1}}}}

\usepackage{xcolor}
\usepackage[T1]{fontenc}

\usepackage{amsfonts}
\usepackage{amssymb}
\usepackage{bm}
\newcommand{\uveci}{\bm{\hat{\mathbf{\imath}}}}
\newcommand{\uvecj}{\bm{\hat{\mathbf{\jmath}}}}
\DeclareRobustCommand{\uvec}[1]{{%
	\ifcsname uvec#1\endcsname
		\csname uvec#1\endcsname
	\else
		\bm{\hat{\mathbf{#1}}}%
	\fi
}}

\begin{document}

\bstctlcite{IEEEexample:BSTcontrol}

% paper title
%\title{(Title in 24-point Times font)}
% If the \LARGE is deleted, the title font defaultfile:///home/mhartman/google_drive/work/LNCMI/Data/2016_03_22/dsp/plots/log/pdf/dsp_0003.pdfs to  24-point.
% Actually, 
% the \LARGE sets the title at 17 pt, which is close enough to 18-point.
%+++++++++++++++++++++++++++++++++++++++++++
\title{\LARGE Characterization of the Vacuum Birefringence Polarimeter at BMV: Dynamical Cavity Mirror Birefringence }
%+++++++++++++++++++++++++++++++++++++++++++
% author names and affiliations
% use a multiple column layout for up to three different
% affiliations
%+++++++++++++++++++++++++++++++++++++++++++
%\author{\authorblockN{J. Clerk Maxwell}
%\authorblockA{School of Electrical and\\Computer Engineering\\
%Somewhere Institute of Technology\\
%City, State 54321--0000\\
%Email: maxwell@curl.edu}
%\and
%\authorblockN{Michael Faraday}
%\authorblockA{(List authors on this line using 12 point Times font\\ - use a second line if necessary)\\
%Microwave Research\\
%City, State/Region, Mail/Zip Code, Country\\
%Email: homer@thesimpsons.com}
%\and
%\authorblockN{Andr\'e M. Amp\`ere \\ }
%\authorblockA{Starfleet Academy\\
%San Francisco, CA 96678-2391\\
%Telephone: (800) 555--1212\\
%Fax: (888) 555--1212}}

%+++++++++++++++++++++++++++++++++++++++++++++++++++

% avoiding spaces at the end of the author lines is not a problem with
% conference papers because we don't use \thanks or \IEEEmembership

% for over three affiliations, or if they all won't fit within the width
% of the page, use this alternative format:
% 
% Another example.
\author{\IEEEauthorblockN{M. T. Hartman\IEEEauthorrefmark{1}, R. Battesti\IEEEauthorrefmark{1}, and C. Rizzo\IEEEauthorrefmark{1}} \\
\IEEEauthorblockA{\IEEEauthorrefmark{1}Laboratoire National des Champs Magn\'etiques Intenses (UPR 3228, CNRS-UPS-UGA-INSA)\\ F-31400 Toulouse Cedex, France\\michael.hartman@lncmi.cnrs.fr}}

% use only for invited papers
%\specialpapernotice{(Invited Paper)}

% make the title area
\maketitle

\begin{abstract}
We present the current status and outlook of the optical characterization of the polarimeter at the Bir\'{e}fringence Magn\'etique du Vide (BMV) experiment.  BMV is a polarimetric search for the QED predicted anisotropy of vacuum in the presence of external electromagnetic fields.  The main challenge faced in this fundamental test is the measurement of polarization ellipticity on the order of $\bm{10^{-15}}$ induced in linearly polarized laser field per pass through a magnetic field having an amplitude and length $\bm{B^{2}L=100\,\mathrm{T}^{2}\mathrm{m}}$.  This challenge is addressed by understanding the noise sources in precision cavity-enhanced polarimetry.  In this paper we discuss the first investigation of dynamical birefringence in the signal-enhancing cavity as a result of cavity mirror motion.
\end{abstract}
%\IEEEoverridecommandlockouts
% no keywords

% For peer review papers, you can put extra information on the cover
% page as needed:
% \begin{center} \bfseries EDICS Category: 3-BBND \end{center}
%
% for peerreview papers, inserts a page break and creates the second title.
% Will be ignored for other modes.
\IEEEpeerreviewmaketitle

% Added this command to remove (gobble!) the page numbers (the correct pages will be added by the IEEE editors, if at all). GG October 2015.
\pagenumbering{gobble}

\section{Introduction}
% no \PARstart
In classical electrodynamics, vacuum is defined as a region where light travels at constant velocity, c.  In a change of definition, the framework of quantum electrodynamics (QED) describes vacuum as a region existing in the lowest energy state.  In QED vacuum, fluctuations in the vacuum energy allow for the spontaneous creation and annihilation of virtual particle-antiparticle pairs.  {Within QED, the Euler-Heisenberg Lagrangian for electromagnetism \cite{Heisenberg1936} is a non-linear electrodynamic theory (NLED) which allows for description of QED vacuum as a polarizable medium \cite{0034-4885-76-1-016401}.}

A measurable consequence of vacuum polarization is vacuum magnetic birefringence (VMB), or the vacuum Cotton-Mouton (cmv) effect \cite{PhysRevD.2.2341}.  A proposed polarimetric scheme to measure this optical anisotropy, $\Delta{n_\mathrm{VMB}}$, in the presence of an applied external magnetic field, $B$, was put forth by Iacopini and Zavattini in 1979 \cite{IACOPINI1979151}.  The principle {is to measure this effect by measuring the polarization ellipticity induced by a phase retardation},
\begin{equation}
	\gamma_\mathrm{VMB} = \frac{2\pi}{\lambda}2L_\mathrm{B}\Delta{n_\mathrm{VMB}} = \frac{2\pi}{\lambda}2L_\mathrm{B}{B^2}{k_\mathrm{VMB}}, \label{eqn:gamma_VMB}
\end{equation}
 between the two polarization states of a linearly polarized laser field of wavelength $\lambda$, probing two passes through a length of anisotropic vacuum, $L_\mathrm{B}$.  The factor,
 \begin{equation}
	k_\mathrm{VMB}= \frac{{3}{\alpha^2}{\hbar^3}}{{15}{\mu_0}{m_\mathrm{e}^4}{c^5}}\approx4\times10^{-24}\,\mathrm{T}^{-2},
\end{equation}
is the QED predicted Cotton-Mouton constant {of vacuum}, where $\alpha$ is the fine-structure constant, $\hbar$ is the reduced Planck constant, $\mu_0$ is vacuum permeability, and $m_\mathrm{e}$ is electron mass.

Astrophysical evidence of VMB has been claimed \cite{doi:10.1093/mnras/stw2798}\cite{Capparelli2017}, however, as a laboratory test of QED, VMB has not yet been directly measured.  The main challenge is the expected level of phase-shift in the probe laser resulting from this minute effect,
 \begin{equation}
	\gamma_\mathrm{VMB}= 5\times 10^{-15}\,\mathrm{rad}\frac{\left[1\,\mathrm{{\mu}{m}}\right]}{\lambda}\frac{B^2 L_\mathrm{B}}{\left[100\,\mathrm{T}^2\mathrm{m}\right]}.
\end{equation}
As a point of reference, leading VMB polarimeters, PVLAS \cite{DellaValle2016} and BMV \cite{doi:10.1063/1.4986871}, have demonstrated differential polarization phase sensitivities on the order of $10^{-12}\,\frac{\mathrm{rad}}{\sqrt{\mathrm{Hz}}}$, and the second generation gravitational-wave interferometer, aLIGO, has shown a differential arm sensitivity of order $10^{-13}\,\frac{\mathrm{rad}}{\sqrt{\mathrm{Hz}}}$ \cite{0264-9381-32-7-074001}.

\section{The BMV Instrument}
\IEEEpubidadjcol
% This is needed anywhere in the 2nd column to make sure the column aligns well
\begin{figure}[h]
  \begin{center}
  \includegraphics[width=\linewidth]{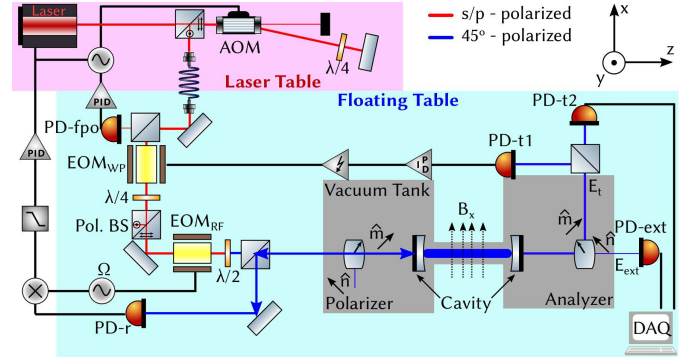}
  \caption[Simplified illustration of the BMV II apparatus.]{Simplified illustration of the BMV II apparatus.  A region of birefringent vacuum is excited by a pulsed magnetic field, $B_\mathrm{x}$, {oriented in the $\uvec{x}$ direction.  The region of birefringent vacuum is probed by a laser field polarized in the $\uvec{m}$ direction ($45^\circ$ between $\uvec{x}$ and $\uvec{y}$) and the resulting phase retardation between the polarization states is enhanced by a resonant cavity.}  The subsequent polarization ellipticity is measured in the extinction channel of the analyzer.}  
  \label{fig:BMVII_diagram_tables}
  \end{center}
\end{figure}

BMV is a cavity-enhanced polarimetric search for vacuum birefringence.  A diagram of the BMV schematic is shown in Fig. \ref{fig:BMVII_diagram_tables}.  The experiment uses a $\lambda = 1\,\mathrm{{\mu}{m}}$ Nd:YAG NPRO laser sent through a single mode fiber from the laser conditioning table to {the main optic table, which is isolated from ground vibration via an active position control system.}  The laser is phase-modulated with the $2^\mathrm{nd}$ {electro-optic modulator ($\mathrm{EOM}_\mathrm{RF}$)} to produce sidebands spaced $\Omega = 10\,\mathrm{MHz}$ from the carrier.  The laser is then polarized in the $\uvec{m}$ direction, $45^\circ$ with respect to the $x$-axis.  The linearly polarized field is incident on a $L_\mathrm{c} = 1.83\,\mathrm{m}$ long high finesse $(\mathcal{F}=440\,000)$ Fabry-Perot cavity.  The carrier frequency is tuned to the cavity resonance, allowing a building up of intracavity field through constructive interference.  The laser field reflected back from the cavity is measured at PD-r and demodulated with $\Omega$ to produce the Pound-Drever-Hall \cite{Drever:83} frequency control-signal, which is fedback to the laser head and {acousto-optic modulator (AOM)} for frequency actuation, keeping the carrier resonant in the cavity.

The apparatus utilizes two power stabilization stages.  The actuation at the AOM produces an unwanted beam deflection; this deflection is reduced by making a double pass back through the AOM with the retro-reflected beam.  Any residual beam position noise is converted to power noise through coupling into a single mode fiber.  This power noise is measured by PD-fpo and suppressed using the AOM as the first power actuator.  Fluctuations of the intracavity power are detected in transmission at PD-t, and suppressed in a second power stabilization loop using the $1^\mathrm{st}$ EOM {($\mathrm{EOM}_\mathrm{WP}$) acting as an electro-optic waveplate in front of a} polarizing beamsplitter. {The bandwidth of the second power stabilization loop is limited by the filtering effect of the light-storage time in the high finesses cavity.  In BMV we have achieved a unity gain frequency of $\approx 40\,\mathrm{kHz}$ with greater than $100 - 1000\times$ suppression of intracavity power noise in the frequency band of $100$ down to $10\,\mathrm{Hz}$}.

{The intracavity field is used to detect birefringent vacuum.}  The $x$ and $y$ polarization components of the linearly polarized laser field experience a different path length due to vacuum birefringence, producing a differential phase between the polarization states.  This differential phase, $\gamma_\mathrm{VMB}$, is enhanced by the cavity finesse,
\begin{equation}
	\Gamma_\mathrm{VMB} = \frac{\mathcal{F}}{\pi}\gamma_\mathrm{VMB},
\end{equation}
amplifying the signal.  The resulting polarization ellipticity is analyzed into a power change by a second polarizer, the `analyzer'.  The signal is ultimately read as the relative power change in the extinction channel of the crossed polarizer,
\begin{equation}
	\frac{\delta{P_\mathrm{ext}}}{\bar{P}_\mathrm{ext}} \approx \frac{2\Gamma_{0}}{4\sigma^2+\Gamma_{0}^2}\Gamma_\mathrm{VMB} \label{eqn-RPN_of_Gamma},
\end{equation}
where $\Gamma_{0}$ is polarization phase delay due to the mean total birefringence of the cavity and $\sigma^2$ is the power extinction ratio of the polarizer-analyzer pair.

%\section{Current Status}

% \begin{figure}
% 	\begin{subfigure}{.25\linewidth}
% 		\centering
% 		\includegraphics[width=.8\linewidth]{xxl-coil_photo_bad}
% 		\caption{1a}
% 		\label{fig:sfig1a}
% 	\end{subfigure}%
% 	\begin{subfigure}{.25\linewidth}
% 		\centering
% 		\includegraphics[width=\linewidth]{Persp_profil_XXL}
% 		\caption{1b}
% 		\label{fig:sfig1b}
% 	\end{subfigure}
% 	\begin{subfigure}{.25\linewidth}
% 		\centering
% 		\includegraphics[width=.8\linewidth]{xxl_temporal}
% 		\caption{1c}
% 		\label{fig:sfig1c}
% 	\end{subfigure}
% \caption{plots of....}
% \label{fig:fig}
% \end{figure}
Commissioning of the second generation experiment, BMV II, is nearing completion. It is clear from \eqref{eqn:gamma_VMB} that the experimental parameter of merit for the vacuum birefringence signal is the product ${B^2}L_\mathrm{B}$.  A significant improvement to BMV II is the upgrade of the signal generating magnet to the new XXL-coil, a pulsed magnet capable of delivering a ${B^2}L_\mathrm{B}$ of up to  $100\,\mathrm{T}^2\mathrm{m}$.  The installation of the new magnet is complete and initial pulses are underway.  The measured rise-time of the pulsed field is $\tau_\mathrm{XXL}=5.9\,\mathrm{ms}$, a value above the characteristic time of the cavity set by the intracavity photon lifetime, $\tau_\mathrm{\gamma}$,
\begin{equation}
	\tau_\mathrm{XXL} > \pi\tau_\mathrm{\gamma} = \pi\frac{\mathcal{F}L_\mathrm{c}}{{\pi}c}\approx 2.7\,\mathrm{ms},
\end{equation}
the condition required to mitigate the signal attenuation resulting from cavity light storage time.  This gives a fundamental frequency of the pulse at $f_\mathrm{XXL} \approx \frac{1}{4\tau_\mathrm{XXL}}\approx 42\,\mathrm{Hz}$.

\begin{figure}[h]
  \begin{center}
  \includegraphics[width=\linewidth]{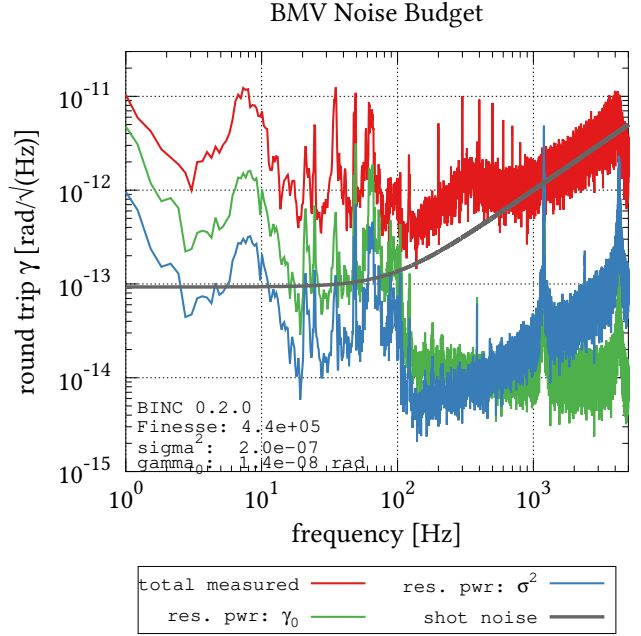}
  \caption[The spectral density of the total measured birefringence fluctuations (\em {red}) compared to select (the largest) modeled sources of sensing noise in the BMV II polarimeter.]{The spectral density of the total measured birefringence fluctuations ({\em red}) compared to select (the largest) modeled sources of sensing noise in the BMV II polarimeter.  Sensing-noise is dominated by shot noise ({\em grey}) at high frequencies.  At low frequencies the largest sensing noise contribution comes from residual laser power noise, coupling through both $\sigma^2$ ({\em blue}) and through the mean birefringence, $\gamma_0$, of the cavity mirrors ({\em green}).  An unknown source of noise limited the polarimeter at frequencies below $500\,\mathrm{Hz}$}
  \label{fig:dsp_0003_lite_sq}
  \end{center}
\end{figure}

This upgrade is done in conjunction with enhancement of the optical setup.  We have developed a thorough set of sensing-noise models \cite{doi:10.1063/1.4986871}, allowing us to understand how fundamental detection noise appears as unwanted intracavity birefringence noise.  This allows us to separate it from the noise due to dynamical cavity birefringence.  An abbreviated noise budget for the BMV II polarimeter for a typical measurement is shown in Fig. \ref{fig:dsp_0003_lite_sq}.  {Here we see the discrepancy at low frequency between the level of actual measured birefringence noise ({\em red}) and the level of sensing noise as predicted by our models (other colors).  This discrepancy indicates a yet unmodeled noise source.  This measured birefringence noise shares features in frequency space with the fluctuations in the intracavity power ({\em green} and {\em blue}), suggesting that the birefringence noise and cavity power noise could have a common source, such as cavity mirror motion.}

There has long been known an inherent birefringence of dielectric cavity mirrors\cite{Bouchiat1982}, observing a birefringence proportional to the square of the angle of incidence on the mirror, $\Gamma_\mathrm{cav}\propto\theta_\mathrm{inc}^2$.  As presented at the 2018 Conference on Precision {Measurements} \cite{8501190}, this paper investigates the effect of small modulations, $\delta\theta(t)$, of this incident angle, $\theta_\mathrm{inc} = \theta_0 + \delta\theta(t)$ on the measured cavity birefringence, observing the linearized term $\delta{\Gamma_\mathrm{cav}}\propto\theta_0\delta\theta$.  Additionally, we discuss this effect as a noise source in the cavity-enhanced polarimeter at BMV.

\section{Methodology}

\begin{figure}[h]
  \begin{center}
  \includegraphics[width=\linewidth]{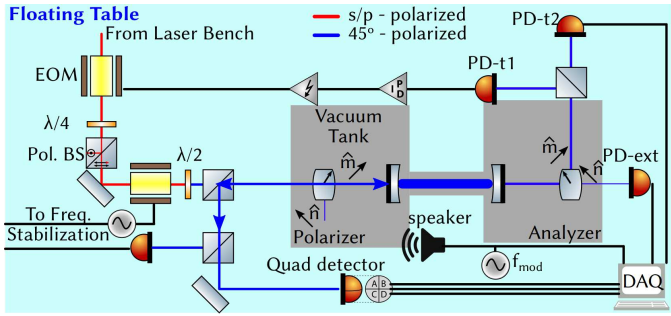}
  \caption[Setup of the cavity mirror modulation measurement.]{Setup of the cavity mirror modulation measurement.  The input mirror is shaken by a loud speaker.  The resulting mirror deflection is measured with a quadrant detector and recorded on the data acquisition card (DAQ) alongside the driving signal and extinction channel photodiode.}  
  \label{fig:mirror_shaker}
  \end{center}
\end{figure}

We measured the effect of cavity mirror motion on the measured cavity birefringence by vibrating the vacuum tank which houses the input mirror, `mirror 1', of the optical cavity by using a loud speaker driven by a sinusoid voltage with frequency $f_\mathrm{mod}$.  A diagram of the measurement is shown in Fig. \ref{fig:mirror_shaker}.  Using in-vacuum accelerometers, care was taken to shake the in-vacuum optics table on the order of the ambient noise of the room, keeping the noise in the same regime.  

The optical signals in transmission of the cavity are split by polarization and analyzed in the usual manner \cite{doi:10.1063/1.4986871} to calculate the birefringence time-series of the cavity, {$\Gamma(t)$}. The angular displacement of mirror 1 is calculated using the lateral displacement of the beam as measured by a quadrant photodetector located at an optical lever $L_\mathrm{mir1,quad} = 1.55\,\mathrm{m}$ in reflection from cavity mirror 1. The quadrant detector measures the displacement parallel and perpendicular to the table ($\uvec{x}$ and $\uvec{y}$).  The cavity mirror is mounted in-vacuum in a PZT driven tip-tilt mount, and to find the principle axes of angular motion ($\uveci$ and $\uvecj$) we excited a resonance of the mount and rotated the coordinates
\begin{equation}
\begin{bmatrix}
\theta_\mathrm{i}\\
\theta_\mathrm{j}
\end{bmatrix}
=
\begin{bmatrix}
\cos{\phi_\mathrm{r}} & -\sin{\phi_\mathrm{r}}\\
\sin{\phi_\mathrm{r}} & \cos{\phi_\mathrm{r}}
\end{bmatrix}
\begin{bmatrix}
\theta_\mathrm{x}\\
\theta_\mathrm{y}
\end{bmatrix}
\end{equation}
of the recorded signals to find the maximum occurring at $\phi_\mathrm{r}=43.6^\circ$, not unexpected as the mount was installed with the tip-tilt axes nominally at $45^\circ$ with respect to the table.

\begin{figure}[h!]
  \begin{center}
  
  \includegraphics[width=\linewidth]{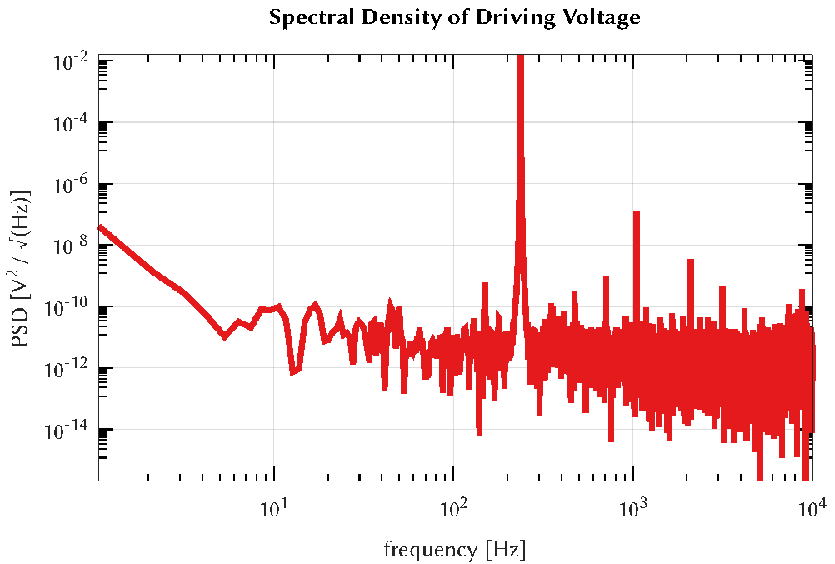}

  \includegraphics[width=\linewidth]{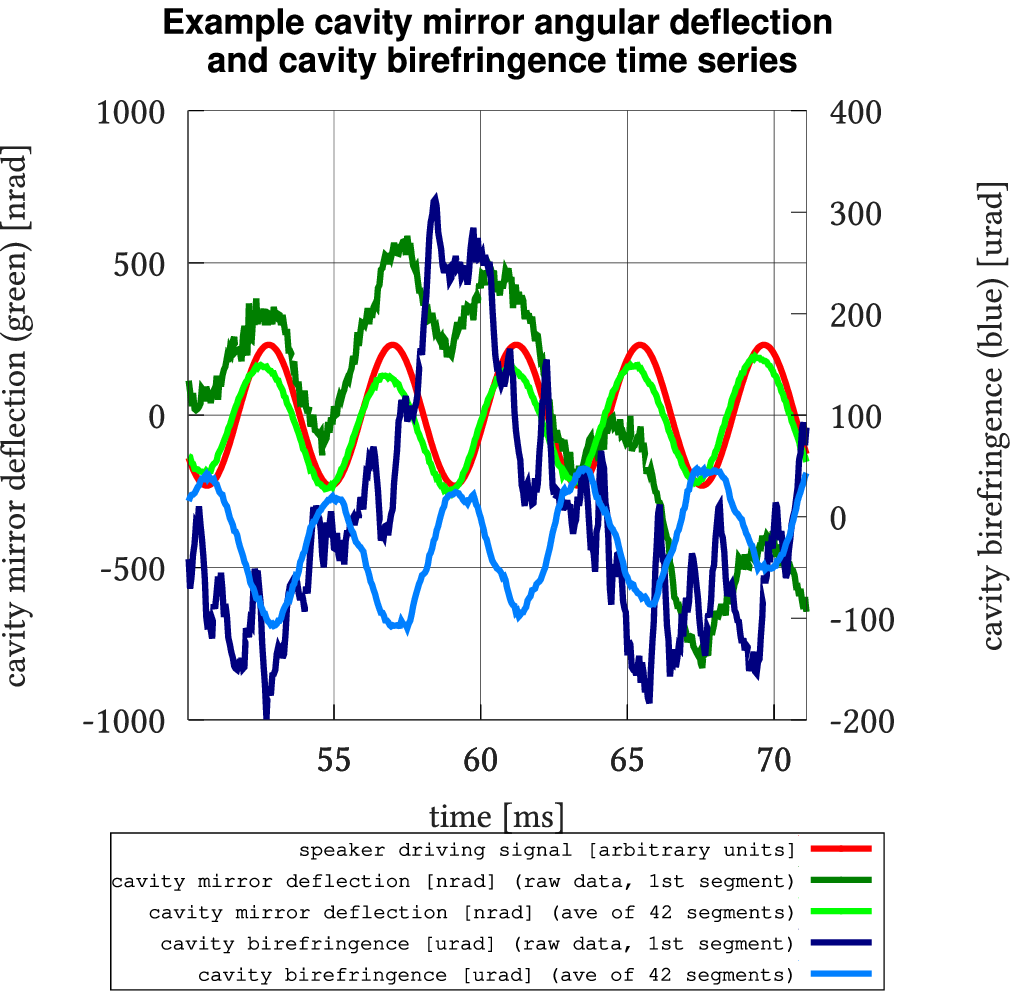}
  
  \caption[An example data point in the mirror modulation to cavity birefringence measurement.]{{An example data point in the mirror modulation to cavity birefringence measurement. {\bf Top:} The PSD of the driving signal is used for the initial guesses in fitting the demodulation sinusoid. {\bf Bottom:} A time-series segment (length: 5 periods of oscillation) for the cavity mirror angular deflection along the $\uveci$ axis, plotted in green.  The {\em dark green} curve shows the first 5-period segment, and the {\em light green} curve shows the average of 42 such segments.  Similarly, the resulting cavity birefringence for the first segment is plotted in {\em dark blue} and the average over 42 segments is shown in {\em light blue}.  For reference, the voltage signal driving the shaker is plotted in {\em red} in arbitrary units. The amplitude and phase of each signal is recovered by I/Q demodulation in post-processing.}}
  \label{fig:step_angDefli2bir_0021}
  \end{center}
\end{figure}
The resulting small birefringence, {$\delta{\Gamma}(t)$}, and mirror angle, $\delta{\theta}(t)$, modulations were recovered using a correlated measure of the sinusoidal voltage which was driving the shaker. {In Fig. \ref{fig:step_angDefli2bir_0021}, we show an example data point plotting the modulation of the cavity mirror birefringence ({\em blue}) resulting from the small (less than background) driven modulation of cavity mirror angle ({\em green})}.  Several segments are averaged {in order to raise the small correlated driven signal above the background birefringence noise. Finally,} the signal amplitudes and phases are recovered via I/Q demodulation using a fitted sinusoid to the driving signal.  The initial guesses for the demodulation sinusoid are calculated using the power spectral density (PSD).  The total power in the driving signal, $P_\mathrm{tot}$, is computed by summing over the power densities, $p_n$, in the individual bins of width $f_\mathrm{bin}$ around the peak bin at $n_\mathrm{pk}$:
\begin{equation}
  P_\mathrm{tot} = f_\mathrm{bin} \sum_{n=n_\mathrm{pk}-3}^{n_\mathrm{pk}+3} p_n
\end{equation}
The amplitude is estimated from the power,
\begin{equation}
  A_\mathrm{est} = \sqrt{2P_\mathrm{tot}},
\end{equation}
and the frequency is estimated by an average of the center frequencies, $f_n$, of the bins around the peak bin, each weighted by their respective power densities:
\begin{equation}
  f_\mathrm{est} = \frac{f_\mathrm{bin}}{P_\mathrm{tot}}\sum_{n=n_\mathrm{pk}-3}^{n_\mathrm{pk}+3} f_n p_n
\end{equation}
{The demodulation sinusoid} is constructed with these parameters and fit in-phase, amplitude and frequency to the driving signal.  The resulting constructed in-phase and quadrature-phase sinusoids act as the I/Q demodulation signals to determine the modulated amplitudes and phases of the cavity mirror angle and cavity birefringence {signals}.

\begin{figure}[h]
  \begin{center}
  \includegraphics[width=\linewidth]{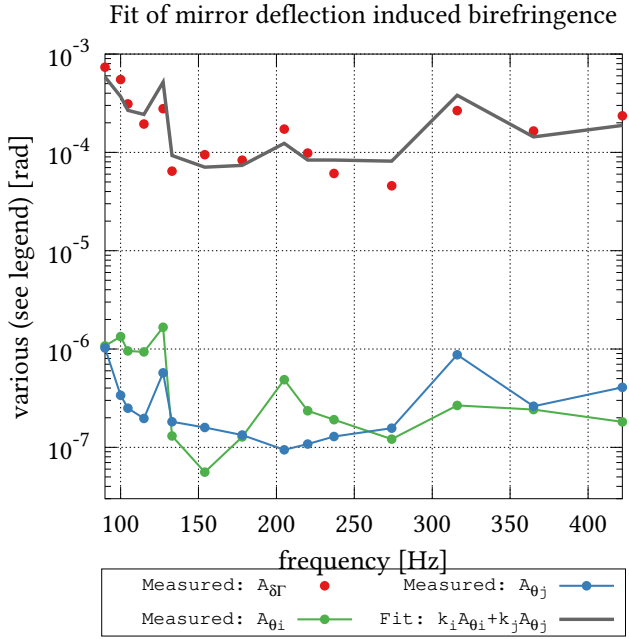}
  \caption[Plot of the modulation amplitudes of each signal for several driving frequencies.]{Plot of the modulation amplitudes of each signal for several driving frequencies.  A sum ({\em grey}) of the angular modulation amplitudes ({\em green} and {\em blue}) is fit to the measured birefringence modulation amplitudes ({\em red})}  
  \label{fig:fit_angle2birefringence}
  \end{center}
\end{figure}

A plot of the angular modulation amplitudes for each axis ($A_\mathrm{\theta{i}}$ ({\em green}) and $A_\mathrm{\theta{j}}$ ({\em blue})) and for each modulation frequency are plotted in Fig. \ref{fig:fit_angle2birefringence} along side the amplitudes of the measured cavity-enhanced birefringence, $A_\mathrm{\delta{\Gamma}}=\frac{\mathcal{F}}{\pi}A_\mathrm{\delta{\gamma}}$, in ({\em red}).  To disentangle the two degrees of freedom, the initial guesses were taken by examining two case frequencies where one axis is excited more than the other: $205\,\mathrm{Hz}$ for $\theta_\mathrm{i}$ and $316\,\mathrm{Hz}$ for $\theta_\mathrm{j}$.  The amplitudes provided an initial guess to the fits, and, for our case, the phases indicated that the angular displacements along the two axes were correlated with each other and both were anti-correlated with the birefringence response.  Armed with these indicators we made a least-squares fit ({\em grey}) of the scaled sum of the angular modulation amplitudes, $k_\mathrm{i}A_\mathrm{\theta{i}} + k_\mathrm{j}A_\mathrm{\theta{j}}$ to the measured birefringence amplitudes. The fitted responses are shown in TABLE \ref{tab:fit_params}.

\begin{table}[h]
\large
\caption{\label{tab:fit_params} The  response coefficients ($k_\mathrm{i}, k_\mathrm{j}$) for both the total measured birefringence ($\frac{\delta{\Gamma}}{\delta{\theta}}$) of the cavity as well as equivalent measured intracavity round-trip birefringence ($\frac{\delta{\gamma}}{\delta{\theta}}$)  produced by modulation of the cavity input mirror for each angular axis.}
\centering
\begin{tabular}{l c c}
 &$\frac{\delta{\Gamma}}{\delta{\theta}}\,\left[\frac{\mathrm{rad}}{\mathrm{rad}}\right]$ &$\frac{\delta{\gamma}}{\delta{\theta}}\,\left[\frac{\mathrm{rad}}{\mathrm{rad}}\right]$\\
\hline
$k_\mathrm{i}$ & $180$ & $7.5\times 10^{-3}$ \\
$k_\mathrm{j}$ & $380$ & $1.6\times 10^{-2}$ \\
\end{tabular}
\end{table}

\section{Discussion}
After the mirror deflection experiment to measure the birefringence response of a cavity to mirror angular deflection, we examine here the potential effect this has on the polarimeter sensitivity in BMV.  A recording of a time-series in unmodulated yet `noisy' conditions for the ellipticity sensitivity was taken to re-examine our noise budget.  These less than ideal conditions are defined by both the literal acoustical noise (air handler in the clean room), and, more importantly, by the characteristics of the cavity installed during this stage in the commissioning of BMV II.  The lower finesse ($\mathcal{F} = 75\,200$) compared to the earlier noise budget ($\mathcal{F} = 440\,000$, Fig. \ref{fig:dsp_0003_lite_sq}) reduces the sensitivity to intracavity birefringence linearly, and the comparatively high static birefringence ($\Gamma_0 = 7\times10^{-2}$ vs. $2\times10^{-3}$) increases the sensitivity to coupling of intracavity power noise to apparent birefringence noise.  

\begin{figure}[h]
  \begin{center}
  \includegraphics[width=\linewidth]{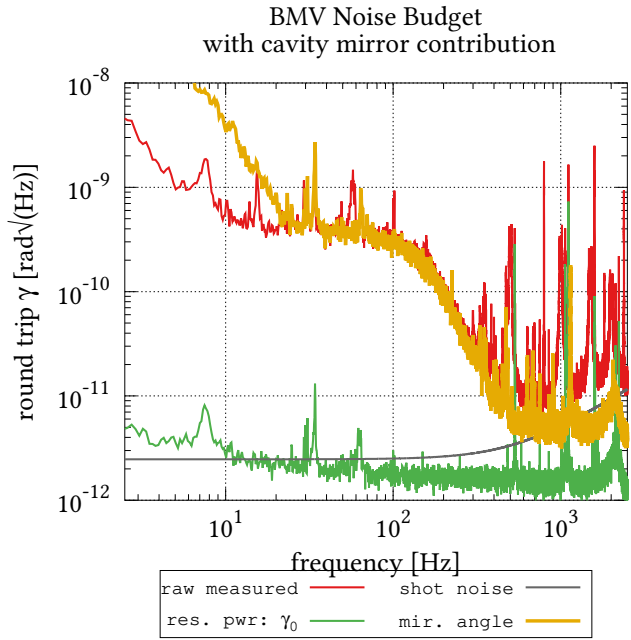}
  \caption[A noise budget for the BMV II polarimeter comparing the spectrum of measured birefringence fluctuations ({\em red}) to select noise sources.]{A noise budget for the BMV II polarimeter comparing the spectrum of measured birefringence fluctuations ({\em red}) to select noise sources.  The newly characterized cavity-mirror angular noise ({\em yellow}) dominates below $500\,\mathrm{Hz}$.  This is a noise budget for a relatively low Finesse cavity with high static birefringence relative to the measurement taken in Fig. \ref{fig:dsp_0003_lite_sq}}  
  \label{fig:noise_budget_with_angle}
  \end{center}
\end{figure}
We take this sensitivity measurement and plot the calculated spectral density of measured raw birefringence sensitivity ({\em red}) on Fig. \ref{fig:noise_budget_with_angle}, along with some of the previously calculated sensing noise models: shot noise in {\em grey} and intracavity power noise coupling through the mean cavity birefringence in {\em green}.  We, of course, add the newly characterized noise source, cavity mirror angular deflection, in {\em yellow}.  This was calculated by scaling the measured angle time-series (as measured by the quadrant detector) by the previously measured fit parameters (TABLE \ref{tab:fit_params}). We then plot the spectral density of the resulting scaled time-series.  The measured magnitude suggests that cavity birefringence noise due to cavity mirror {angular motion} is a limiting noise source for this sensitivity run, matching the magnitude of the measured birefringence noise in the frequency band from $20\,\mathrm{Hz}$ to $500\,\mathrm{Hz}$. 

There are notable peaks (ex. near $30$ and $57\,\mathrm{Hz}$) which {are} unexplained by the measured mirror birefringence noise.  Possible sources for the remaining peaks include the possibility {that} they {are} the result of {angular motion} in the second (unmeasured) cavity end-mirror, or perhaps they have a different, still unknown, source altogether. Additionally, the calculated noise contribution from angular motion exceeds the measured birefringence noise at frequencies below $20\,\mathrm{Hz}$.  This could be the result of a beam motion on the quad-detector which is not due to angular motion of the cavity mirror.  Alternatively, a change in the magnitude of the response at low-frequencies (below our measurement band), or a change in the phase relationship between the mirror mount axes and the resulting birefringence would cause them to loose coherence at low frequencies, decreasing the effective response.  These possibilities can be investigated with additional equipment to measure and record beam motion in the injected beam and the motion of the cavity end mirror.

\begin{figure}[h]
  \begin{center}
  \includegraphics[width=\linewidth]{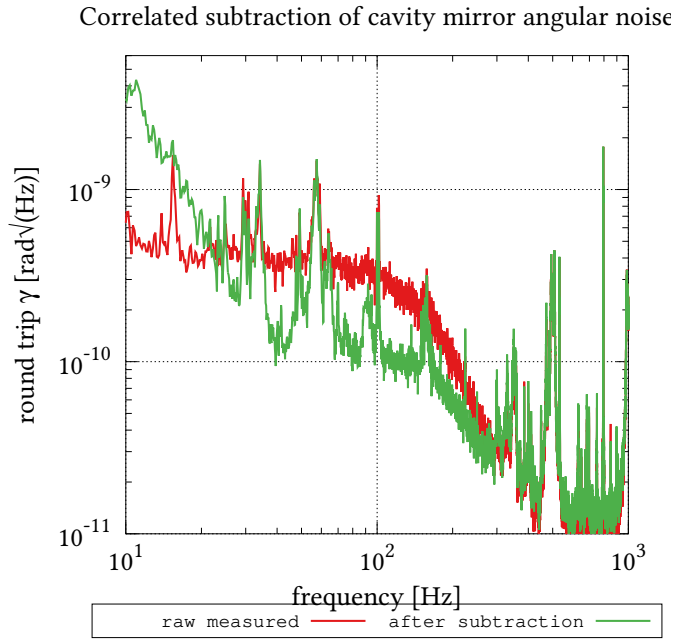}
  \caption[Spectral density of the birefringence time-series after coherent subtraction of cavity mirror {angular motion time series} ({\em green}).]{Spectral density of the birefringence time-series after coherent subtraction of cavity mirror {angular motion time series} ({\em green}).  The raw birefringence signal, the same in Fig. \ref{fig:noise_budget_with_angle}, is plotted in {\em red} in both Figures.}  
  \label{fig:noise_subtracted}
  \end{center}
\end{figure}

Finally, we test the correlation of the mirror angular noise measurement by subtracting the {scaled angular motion} time-series from the raw measured birefringence time-series to attempt a coherent subtraction of this noise source:
\begin{equation}
	\delta{\gamma_\mathrm{res}}(t) = \delta{\gamma_\mathrm{raw}}(t)- k_\mathrm{i}\delta{\theta_\mathrm{i}}(t) - k_\mathrm{j}\delta{\theta_\mathrm{j}}(t)
\end{equation}
We then take the power spectral density of the residual noise time-series and compare the resulting curve ({\em green}) with the raw birefringence spectrum ({\em red}) in Fig. \ref{fig:noise_subtracted}.  The result is a $5\times$ reduction in noise for much of the band from $30-200\,\mathrm{Hz}$ including at 42 Hz, currently the largest Fourier component of the signal generating magnet.

A measurement of the birefringent response of the cavity to mirror motion allows us to discuss the vacuum-birefringence based requirements for cavity mirror stability.  With an expected differential phase retardation signal on the order of $\tilde{\gamma}_\mathrm{VMB} \approx 10^{-15}\,\frac{\mathrm{rad}}{\sqrt{\mathrm{Hz}}}$ we can naively propose a requirement of mirror angular stability of $\tilde{\theta}_\mathrm{req} = \frac{\tilde{\gamma}_\mathrm{VMB}}{k_\mathrm{j}}\approx 0.1\,\frac{\mathrm{prad}}{\sqrt{\mathrm{Hz}}}$.  This is an incomplete story, however, as this assumes the effect is the cavity-amplified birefringence of the angle between the reflective coating of the cavity mirror and the intracavity beam.  Alternatively, the effect could be the angle between the incident beam and the input of the cavity mirror.  While not amplified by the cavity's phase response, this beam sees more optical material and could be more significant in a low-finesse case.  Additionally, the magnitude of the response to mirror angular noise could depend on the alignment of the mirror's optical axis to the polarization of the laser field.  These considerations can be investigated using cavities of varying finesse and optical alignment.

\section{Conclusion and Outlook}

The correlated measurement of cavity mirror angle with cavity birefringence for the purpose of noise subtraction in VMB searches is complicated by the phase shifts through numerous mechanical resonances of system, in particular the cavity mirror mount.  Additionally, the magnitude of the effect should depend on the position of the mirror sampled by the beam and the microscopic offset angle,  $\theta_0$, the laser makes with the normal of the mirror.  Therefore, the angular cavity mirror birefringence effect would change with the small optical misalignments evolving in time, making noise subtraction difficult over multi-day pulsed magnet measurement campaigns.

The identification of this noise source and its order of magnitude, however, does allow us to discuss vacuum birefringence-based sensitivity requirements on the allowed angular motion of the cavity mirrors.  As a result, in addition to real-time correlated noise subtraction, BMV turns its attention to the mitigation of mirror motion induced birefringence through vibration reduction in its steps towards future generation detectors.  The more immediate steps will involve a study of this effect in the high-finesse cavity that BMV will use in its upcoming measurement campaign, where one can imagine the stricter alignment requirements of the high finesse reduces the allowed $\theta_0$, subsequently changing the relative values of the $\theta_0\delta{\theta}$ and $\delta{\theta}^2$ terms.

Finally, with the installation work largely completed, we are currently characterizing the pulsed magnetic {field} in the polarimeter.  This is done by measuring systematic effects correlated with the pulse of the magnetic field.  The system will be calibrated by measuring the Cotton-Mouton effect of well known gases, nitrogen and helium \cite{PhysRevA.88.043815}, the later providing a test of sensitivity as it has the smallest known Cotton-Mouton effect of $2.4\times10^{-16}\,\mathrm{T}^{-2}\mathrm{atm}^{-1}$.  After calibration, commissioning will continue with vacuum measurements and further work on noise reduction.

% conference papers do not normally have an appendix

\section*{Acknowledgment}
This research has been partially supported by ANR (Grant No. ANR-14-CE32-0006) of the framework of the ``Programme des Investissements d'Avenir''.  We thank the members of the BMV collaboration.

% optional entry into table of contents (if used)
%\addcontentsline{toc}{section}{Acknowledgment}

% trigger a \newpage just before the given reference
% number - used to balance the columns on the last page
% adjust value as needed - may need to be readjusted if
% the document is modified later
%\IEEEtriggeratref{8}
% The "triggered" command can be changed if desired:
%\IEEEtriggercmd{\enlargethispage{-5in}}

% references section
% NOTE: BibTeX documentation can be easily obtained at:
% http://www.ctan.org/tex-archive/biblio/bibtex/contrib/doc/

% can use a bibliography generated by BibTeX as a .bbl file
% standard IEEE bibliography style from:
% http://www.ctan.org/tex-archive/macros/latex/contrib/supported/IEEEtran/bibtex
%\bibliographystyle{IEEEtran.bst}
% argument is your BibTeX string definitions and bibliography database(s)
%\bibliography{IEEEabrv,../bib/paper}
%
% <OR> manually copy in the resultant .bbl file
% set second argument of \begin to the number of references
% (used to reserve space for the reference number labels box)

\bibliographystyle{IEEEtran.bst}
\bibliography{IEEEabrv,\string~/google_drive/work/my_references.bib}

% \begin{thebibliography}{1}
% 
% \bibitem {Bierzychudek}
% M. E. Bierzychudek and R. E. Elmquist, “Uncertainty evaluation in a two-terminal cryogenic current comparator,” \emph{IEEE Trans. Instrum. Meas.}, vol. 58, no. 4, pp. 1170 – 1175, April 2009.
% 
% \bibitem {Jarret}
% D. G. Jarrett and R. E. Elmquist, ``Settling time of high-value standard resistors,'' CPEM 2004 Conf. Digest, p. 522, June 2004.
% 
% \bibitem {krauss}
% H. L. Krauss, C. W. Bostian, and F. H. Raab, \emph{Solid State Radio Engineering}, New York: J. Wiley \& Sons, 1980.
% 
% %\bibitem{IEEEhowto:kopka}
% %H.~Kopka and P.~W. Daly, \emph{A Guide to {\LaTeX}}, 3rd~ed.\hskip 1em plus
% % 0.5em minus 0.4em\relax Harlow, England: Addison-Wesley, 1999.
% 
% %\bibitem{lamport} L. Lamport, \emph{ {\LaTeX} A Document Preparation
% %  System}, Reading, Mass: Addison-Wesley, 1994.
% 
% %\bibitem{knuth} D. E. Knuth, \emph {The \TeX book}, Reading, Mass.:
% %  Addison-Wesley, 1996.
% 
% \end{thebibliography}

% that's all folks

\end{document}